\documentclass[sigconf]{acmart}

\copyrightyear{2024}
\acmYear{2024}
\setcopyright{rightsretained}
\acmConference[UbiComp Companion '24]{Companion of the 2024 ACM International Joint Conference on Pervasive and Ubiquitous Computing }{October 5--9, 2024}{Melbourne, VIC, Australia}
\acmBooktitle{Companion of the 2024 ACM International Joint Conference on Pervasive and Ubiquitous Computing (UbiComp Companion '24), October 5--9, 2024, Melbourne, VIC, Australia}
\acmDOI{10.1145/3675094.3677555}
\acmISBN{979-8-4007-1058-2/24/10}

\usepackage{graphicx}
\usepackage{subfig}
\usepackage{enumitem}

\AtBeginDocument{%
  }

\usepackage{soul}
\begin{document}

\title{Impact of Spatial Proximity on Drone Services}

\author{Vejaykarthy Srithar}
\affiliation{%
  \institution{The University of Sydney}
  \city{Sydney, New South Wales}
  \country{Australia}}
\email{vsri6304@uni.sydney.edu.au}
\author{Syeda Amna Rizvi}
\affiliation{%
  \institution{The University of Sydney}
  \city{Sydney, New South Wales}
  \country{Australia}}
\email{amna.rizvi@sydney.edu.au}
\author{Amani Abusafia}
\affiliation{%
  \institution{The University of Sydney}
  \city{Sydney, New South Wales}
  \country{Australia}}
\email{amani.abusafia@sydney.edu.au}
\author{Athman Bouguettaya}
\affiliation{%
  \institution{The University of Sydney}
  \city{Sydney, New South Wales}
  \country{Australia}}
\email{athman.bouguettaya@sydney.edu.au}
\author{Balsam Alkouz}
\affiliation{%
  \institution{The University of Sydney}
  \city{Sydney, New South Wales}
  \country{Australia}}
\email{balsam.alkouz@sydney.edu.au}

\renewcommand{\shortauthors}{Vejaykarthy Srithar, Syeda Amna Rizvi, Amani Abusafia, Athman Bouguettaya, \& Balsam Alkouz}

\begin{abstract}
We demonstrate the peer-to-peer impact of drones flying in close proximity. Understanding these impacts is crucial for planning efficient drone delivery services. In this regard, we conducted a set of experiments using drones at varying positions in a 3D space under different wind conditions. We collected data on drone energy consumption traveling in a skyway segment. We developed a Graphical User Interface (GUI) that plots drone trajectories within a segment. The GUI facilitates analyzing the peer-to-peer influence of drones on their energy consumption. The analysis includes drones' positions, distance of separation, and wind impact. \end{abstract}



\maketitle

\section{Introduction}

The proliferation of drone-based delivery services reflects the increasing integration of drones in urban areas. A SESAR study in 2016 predicts 100,000 drones in Europe by 2050 \cite{european}. Recent studies leverage the service paradigm to model drones-as-a-service \cite{shahzaad2019composing}. These drone services usually operate in a skyway network,i.e., an interconnected set of nodes. The nodes are building rooftops that serve as recharging stations or delivery destinations. Drones may recharge at nodes for long-mile deliveries due to limited battery capacity. These nodes are connected through skyway segments. A skyway segment is an aerial space that extends between two adjacent rooftop buildings. It provides a Line of Sight (LoS) path for drones to transit from one rooftop to another.\looseness=-1

Recent studies focus on determining the optimal path  for drones in a skyway network \cite{shahzaad2019composing}. This optimal path is defined by the set of segments that enable efficient package delivery. However, the growing demand for drone deliveries strains the limited airspace in skyway segments \cite{labib2021rise}. As a result, Drones operating in close proximity may \textit{impact} each other. For example, the downwash forces from a higher drone can affect the energy consumption of a lower one \cite{alkouz2020formation}. As a result, drones may deviate from their optimal path, causing a significant risk of \emph{collision}.  Additionally, it may delay package delivery due to increased recharging needs from the additional strain on propellers and batteries. Hence, it is paramount to study the impact of drones on each other's energy.

Existing research studied the inter-impact of a drone swarms in terms of energy consumption \cite{alkouz2020formation}. They presented the energy consumption of each drone in a swarm in different formations, (vee, echelon, column, front). However, they did not consider their peer-to-peer positions  in a 3D space, i.e., vertical, horizontal, and diagonal.  \textit{To the best of our knowledge, there is no analysis of the impact of drones on each other’s energy consumption in varying peer-to-peer positions.} This paper is the first to analyze the impact of spatial proximity on drone energy consumption. We conducted experiments to record battery consumption of drones travelling: (1) individually in a segment and (2) in the presence of other drones in the same segment, at varying distances and under different wind conditions. Finally, we present a Graphical User Interface (GUI) that plots the drones' trajectories in a segment. The GUI helps analyze the impact of varying spatial distances on the drones' battery consumption in the presence of another drone.

\vspace{-5pt}
\section{Overview}
This section describes our experimental setup, data collection process, and platform interface.

\vspace{-5pt}
\subsection{Indoor Drone Testbed}
\begin{figure}\vspace{-10pt}
    \centering
    \includegraphics[width=0.6\linewidth]{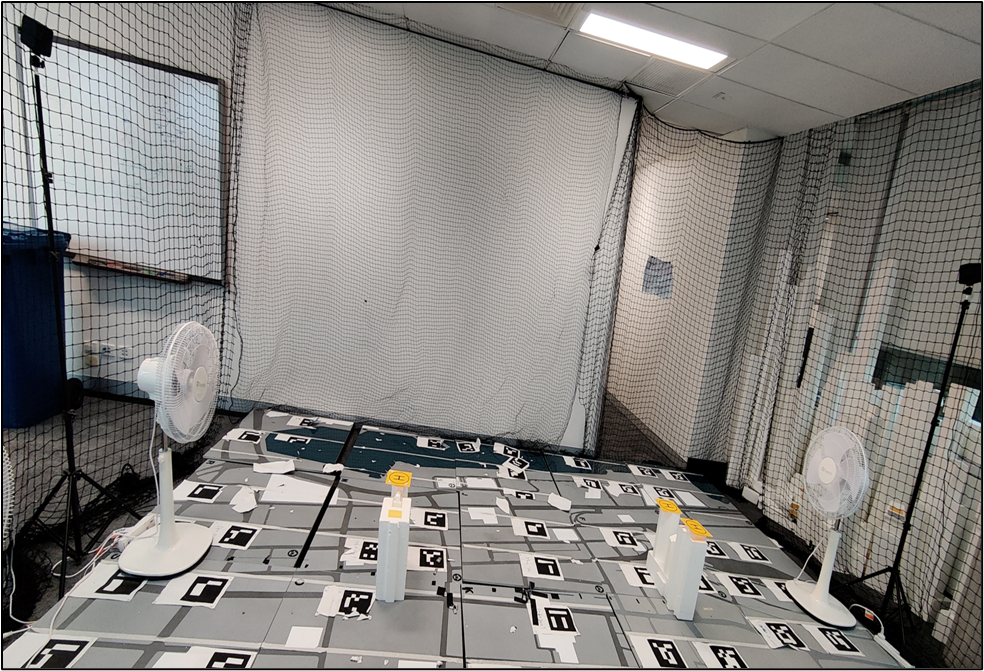}
    \setlength{\abovecaptionskip}{1pt}
    \setlength{\belowcaptionskip}{-15pt}
    \caption{Indoor Testbed}
    \label{fig:Indoor testbed}
\end{figure}

The outdoor environment is unsuitable for experimental purposes due to safety concerns. The government regulations restrict flying drones in certain regions \cite{jones2017international}. An indoor testbed is used to conduct the experiments. The testbed is a small-scale model of dimensions 4m x 5m x 1.7m that resembles Sydney CBD. For simplicity, we consider only two nodes, which serve as source and destination nodes. These nodes are the 3D-printed models of buildings deployed with two recharging pads each. The length of the skyway segment between the source and destination node is 1.4 meters (See Fig. \ref{fig:Indoor testbed}). We used two Crazyflie 2.1 nano-quadcopter drones\footnote{https://www.bitcraze.io/products/crazyflie-2-1/}. We used two HTC Vive V1 base stations that emit infrared lasers. These lasers are received by the Lighthouse Positioning System decks mounted on the drones, allowing accurate positioning within the experiment environment. The wind was generated by a FANCO DC pedestal fan with model number PPFC40DCWH.\looseness=-1
\begin{figure}
    \centering
    \includegraphics[width=0.5\linewidth, height=0.6\linewidth]{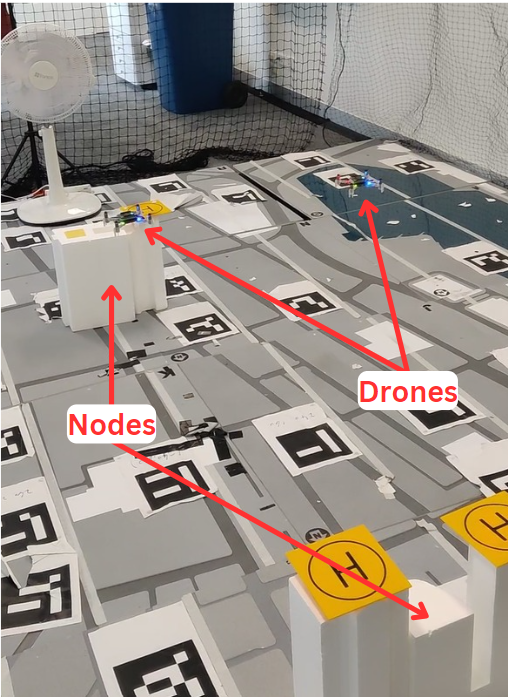}
    \setlength{\abovecaptionskip}{1pt}
    \setlength{\belowcaptionskip}{-18pt}
    \caption{Two Crazyflie Drones in Flight}
    \label{fig:Two Crazyflie drones in flight}
\end{figure}
\vspace{-8pt}
\subsection{Experiments and Data Collection}
We ran several experiments in the indoor drone testbed to understand the impact of drones on each other. The experiments varied in \emph{settings}, \emph{positions}, and \emph{proximities} to other drones. We collected various drone parameters using a Python script to assess the impact on \emph{battery consumption rates}. In this regard, we gather drones' battery levels when operating individually and in the presence of other drones in the same segment. Additionally, we record their X, Y, and Z coordinates, roll, pitch, and yaw to track the stability of the drones' trajectory. These parameters are collected for varying peer-to-peer spatial positions under different wind conditions and separation distances. We consider various separation distances (i.e., 0.3m, 0.4m, 0.5m, 0.6m, 0.7m,.0.8m, and 1m) along the x,y, and z axes for the spatial positions. The initial battery voltage for the drones was around 4.1 Volts. We conducted 224 trips of drone flights.

We used Crazyflie’s Python APIs to communicate with the drones.  
The drones operate at varying distances from the source to the destination node based on input coordinates and speed. 

We then record the peer-to-peer impact on battery consumption when they operate in varying spatial proximities. We also ran the experiments in controlled (i.e., different wind speeds and directions) and uncontrolled (i.e., no wind) conditions.\looseness=-1  

\subsubsection{\textbf{Spatial Positions Under Uncontrolled Conditions}} 
We conducted this set of experiments under uncontrolled wind conditions (i.e., no wind). In each experiment, the drone's battery consumption is recorded when it operates from the source to the destination node.  We started by recording the drone flying alone from source to destination.  We then repeated the experiment with two drones in five different spatial positions. This allowed us to analyze the peer-to-peer impact on drones' battery consumption. We also repeated the same five experiments while increasing the distance of separation between drones. Below we describe all possible peer-to-peer spatial positions of drones in a skyway segment.\looseness=-1

\begin{figure}[!t]
 \vspace{-10pt}
    \centering
\includegraphics[width=0.92\linewidth,height=0.23\textheight]{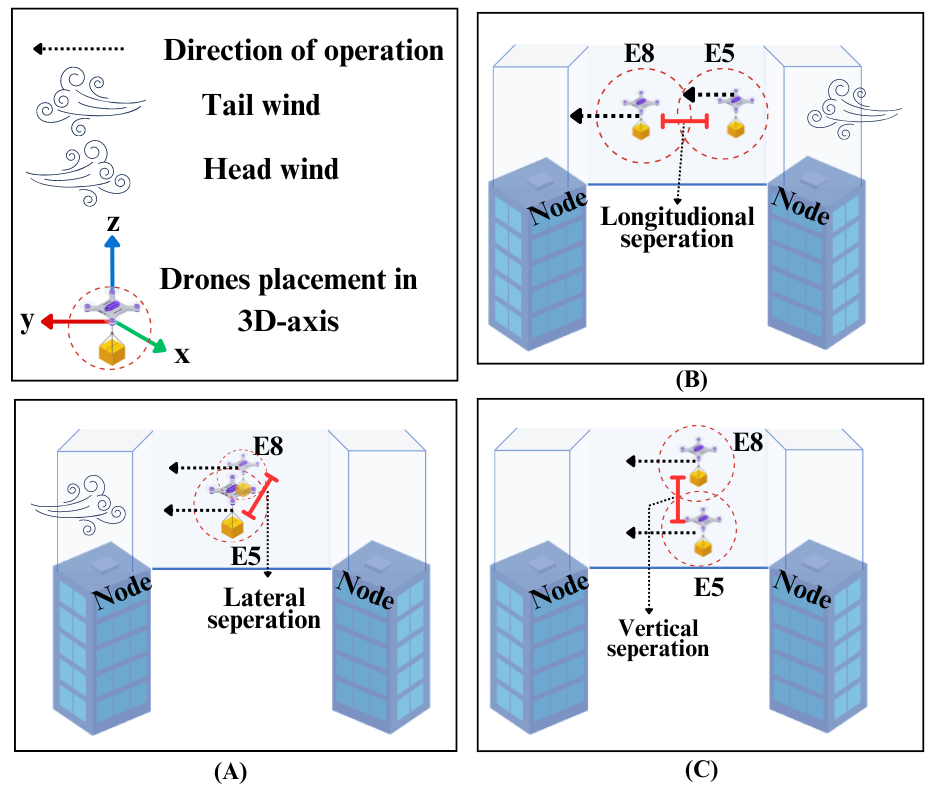}

    \setlength{\abovecaptionskip}{1pt}
    \setlength{\belowcaptionskip}{-15pt}
    \caption{Different Types of Spatial Positioning of Drones}
    \label{Dist}
\end{figure}

\begin{enumerate}[ noitemsep,nosep,leftmargin=*,labelsep=2pt,itemindent=0pt, labelwidth=1pt]
\item \textbf{Side-by-Side:}  The drones are operating at the same altitude (i.e., z-axis) along the y-axis but have \emph{lateral separation} along the x-axis (see Fig. \ref{Dist}(A) with drone E5 is on the left of drone E8).\looseness=-1
\item \textbf{Front-Back:} The two drones operate at the same z-axis along the y-axis and have \emph{longitudinal separation} along the x-axis (see Fig. \ref{Dist}(B) with drone E5 in front and E8 behind). 

\item \textbf{Top-Down:} One drone operates at a higher altitude (i.e., z-axis) than the other along the same y-axis (see Fig. \ref{Dist}(C) where drone E5 operates on the bottom and E8 operates on the top). 

\item \textbf{Side-by-Side with Offset:} The drones fly at the same altitude and are separated along the x-axis. One drone is leading ahead along the x-axis. The other drone is at a fixed lateral offset (i.e., along the y-axis) with the leading drone positioned behind.\looseness=-1 

\item \textbf{Top-Down with Offset:}
One drone is on top, and the other is at the bottom. The drones fly along the same line on the y-axis but are separated along the z-axis with a fixed offset along the y-axis. 
\end{enumerate}
\vspace{-3pt}

\subsubsection{\textbf{Spatial Positions Under Controlled Conditions}}In this set of experiments, we analyze the impact on peer-to-peer battery consumption of drones under controlled conditions (i.e., different wind conditions). In this regard, we consider the aforementioned spatial positions subject to four types of extrinsic wind states: high-speed headwind, low-speed headwind, high-speed tailwind, and low-speed tailwind. The fan is placed behind the destination node facing the source node for headwind direction and behind the source node facing the destination node for tailwind direction. The speed settings are 7.6 km/hr for high and 6.1 km/hr for low, as higher wind speeds caused the drones to become very unstable.

\vspace{-8pt}
\subsection{Graphical User Interface}
\begin{figure}[!t]
 \vspace{-10pt}
    \centering   \includegraphics[width=0.9\linewidth]{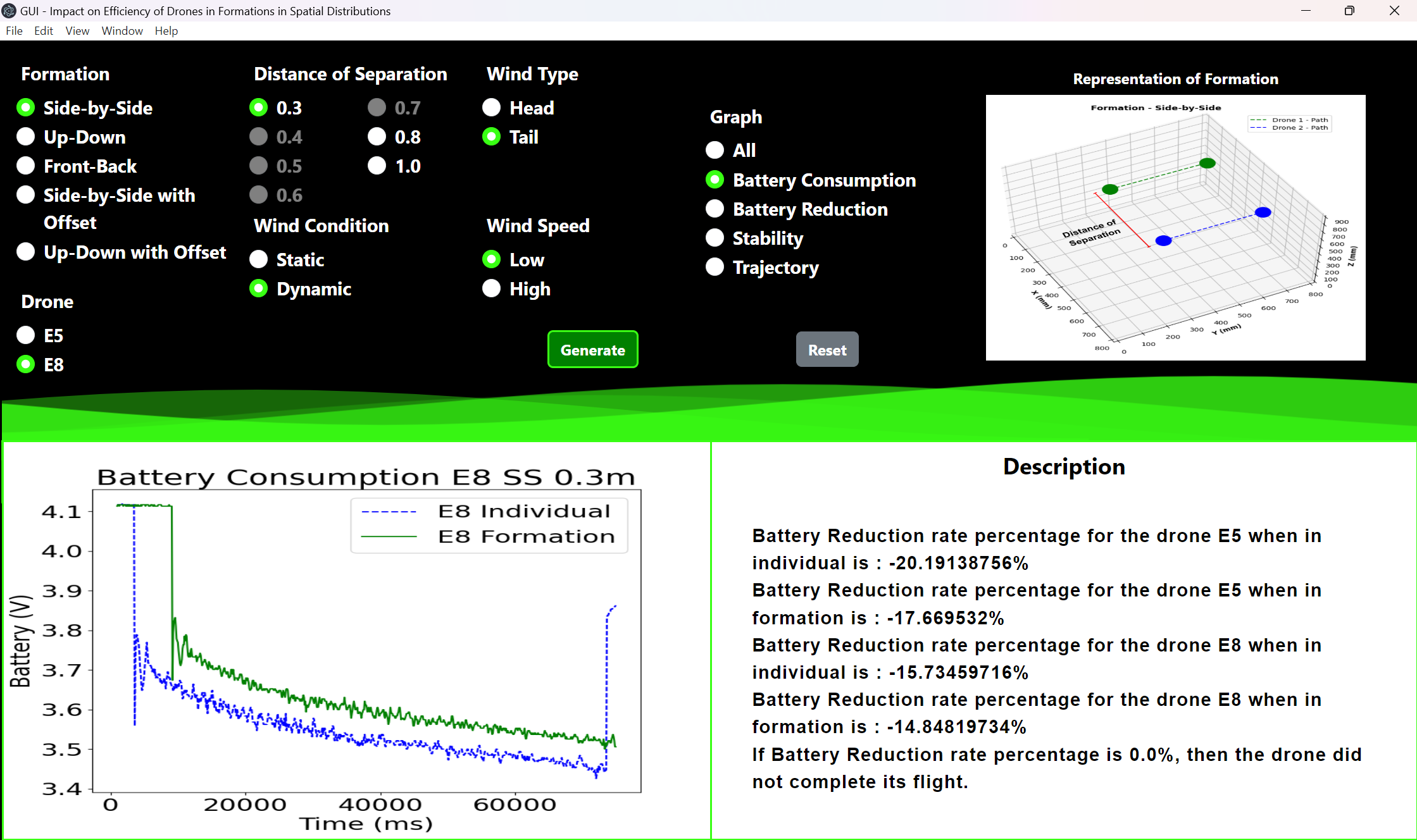}
        \setlength{\abovecaptionskip}{1pt}
        \setlength{\belowcaptionskip}{-10pt}
    
    \caption{GUI screenshot}
    \label{fig:GUI}
    \vspace{-5pt}
\end{figure}

 We developed a GUI using ElectronJS to display the plots of drones' battery consumption drones when they operate individually and in the presence of other drones (See Fig.\ref{fig:GUI}). It takes user inputs for peer-to-peer spatial positions, separation distance, wind direction, and speed. This GUI functions as a facilitator by eliminating the need for manual parsing of extensive CSV files. It also provides concise summaries of peer-to-peer impact on the battery consumption of drones across diverse conditions based on the user's selection.

\section{Results}
The analysis of the collected data involved plotting battery reduction percentage, battery consumption, trajectories, roll, pitch, and yaw. Our focus is on the battery consumption of the drones when they fly in different spatial positions.  The results show that in most cases there is an \emph{increase in the battery consumption of drones due to peer-to-peer impact when they operate nearby}.  Below, we present a detailed analysis of peer-to-peer impact for each spatial position considering different wind conditions.\looseness=-1

\vspace{-8pt}
\subsection{Impact Under Uncontrolled Conditions}
In uncontrolled conditions (i.e., no wind), the battery consumption of both drones is negatively impacted for all spatial positions and separation distances.
The impact starts at approximately less than 0.5m distance between drones for both side-by-side and side-by-side with offset spatial positions. The impact decreases as the distance of separation increases for both front-back and top-down with offset spatial positioning. As shown in Fig. \ref{fig:Static-FB-1}, the drone E5 consumes more battery when it operates near drone E8 than when it operates individually. However, as the longitudinal separation between E8 and E5 increases, the drone at the back consumes less battery (see Fig.  \ref{fig:Static-FB-3}). It is mainly due to the lift by the upwash forces from drone E8 flying in front.\looseness=-1

\vspace{-3pt}
\subsection{Impact Under Controlled Conditions}
In controlled conditions (i.e., different wind speeds and directions), drones' battery consumption is affected by both spatial proximity and external wind. We discuss the most prominent observations in each spatial positioning.

\begin{figure}[!t]
\vspace{-20pt}
  \centering
  \subfloat[Distance of Separation - 0.3m]{\includegraphics[width=0.23\textwidth]{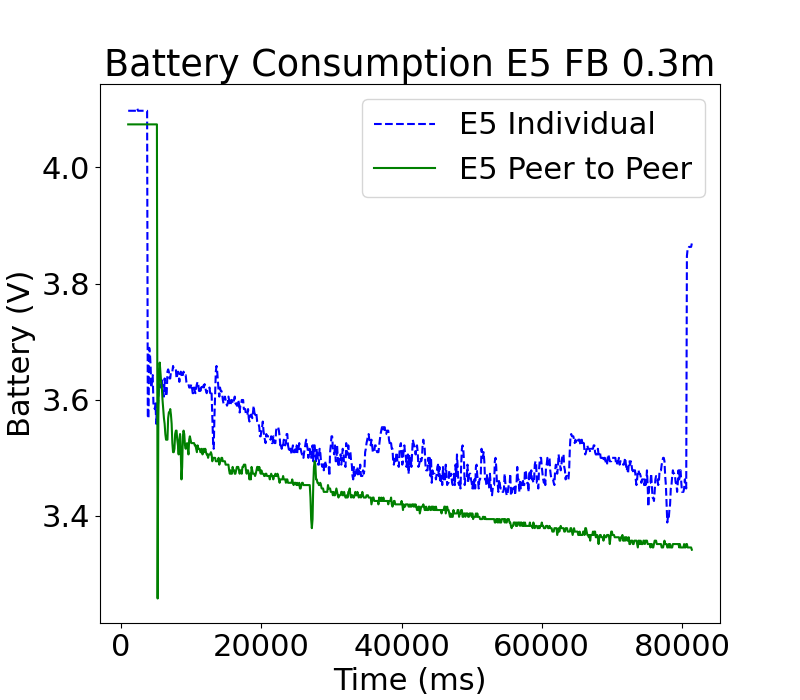}\label{fig:Static-FB-1}}
  \hfill
  \subfloat[Distance of Separation - 0.5m]{\includegraphics[width=0.23\textwidth]{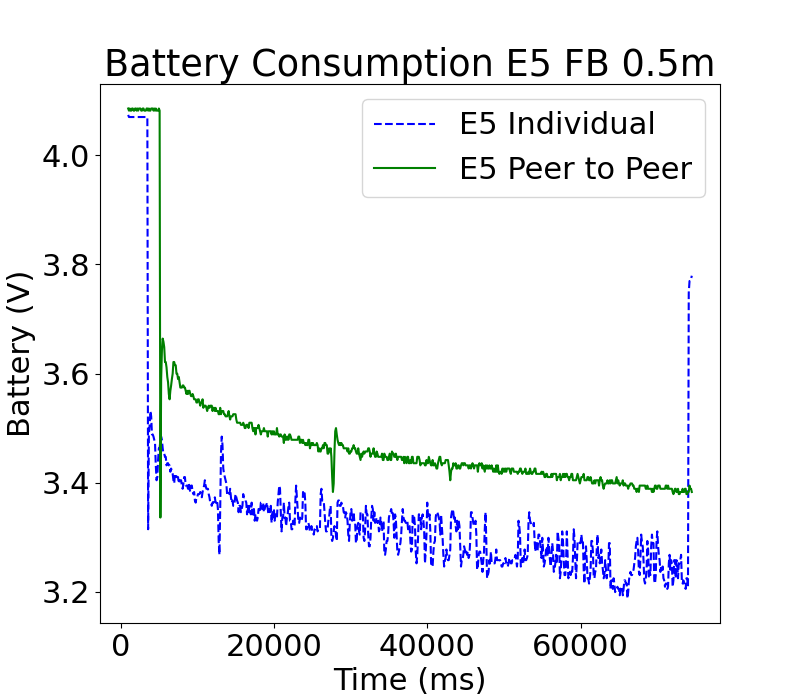}\label{fig:Static-FB-3}}
    \setlength{\abovecaptionskip}{1pt}
    \setlength{\belowcaptionskip}{-15pt}
  \caption{Battery consumption of E5 in different separation distances in front-back spatial position without wind}
  \label{Static-FB}
\end{figure}

\begin{figure}[!t]
\vspace{-8pt}
  \centering
  \subfloat[Distance of Separation-0.3m]{\includegraphics[width=0.23\textwidth]{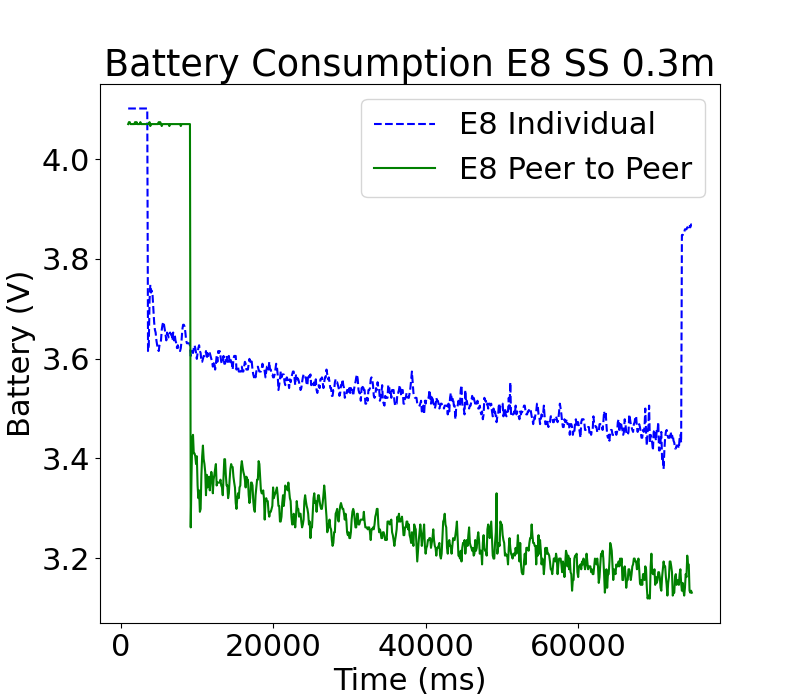}\label{fig:Dynamic-SS-1}}
  \hfill
  \subfloat[Distance of Separation-0.6m]{\includegraphics[width=0.23\textwidth]{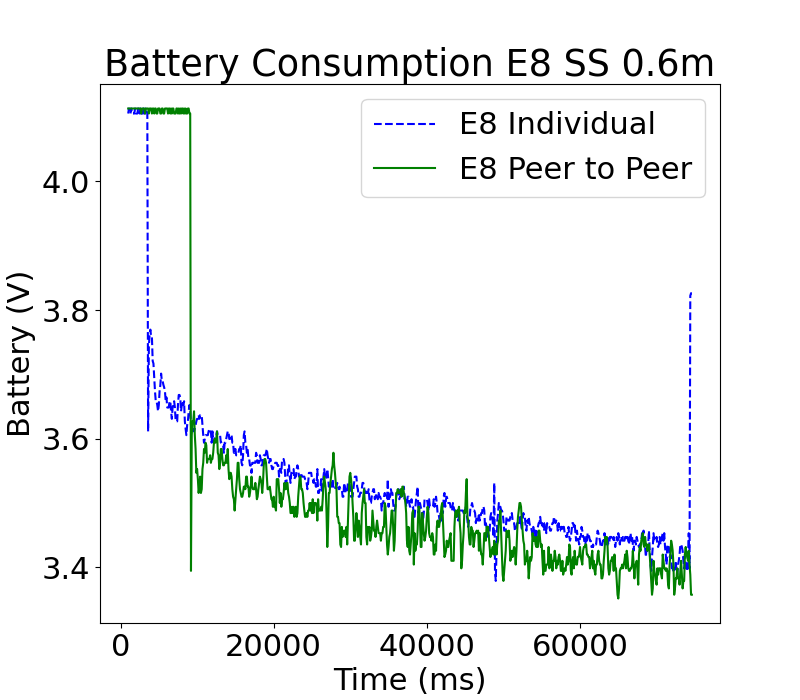}\label{fig:Dynamic-SS-3}}
  \setlength{\abovecaptionskip}{1pt}
    \setlength{\belowcaptionskip}{-15pt}
  \caption{Battery consumption of E8 operating in side-by-side under high-speed headwind}
\end{figure}

\begin{figure}[!b]
\vspace{-18pt}
  \centering
  \subfloat[Drone E5 - Unstable]{\includegraphics[width=0.8\linewidth]{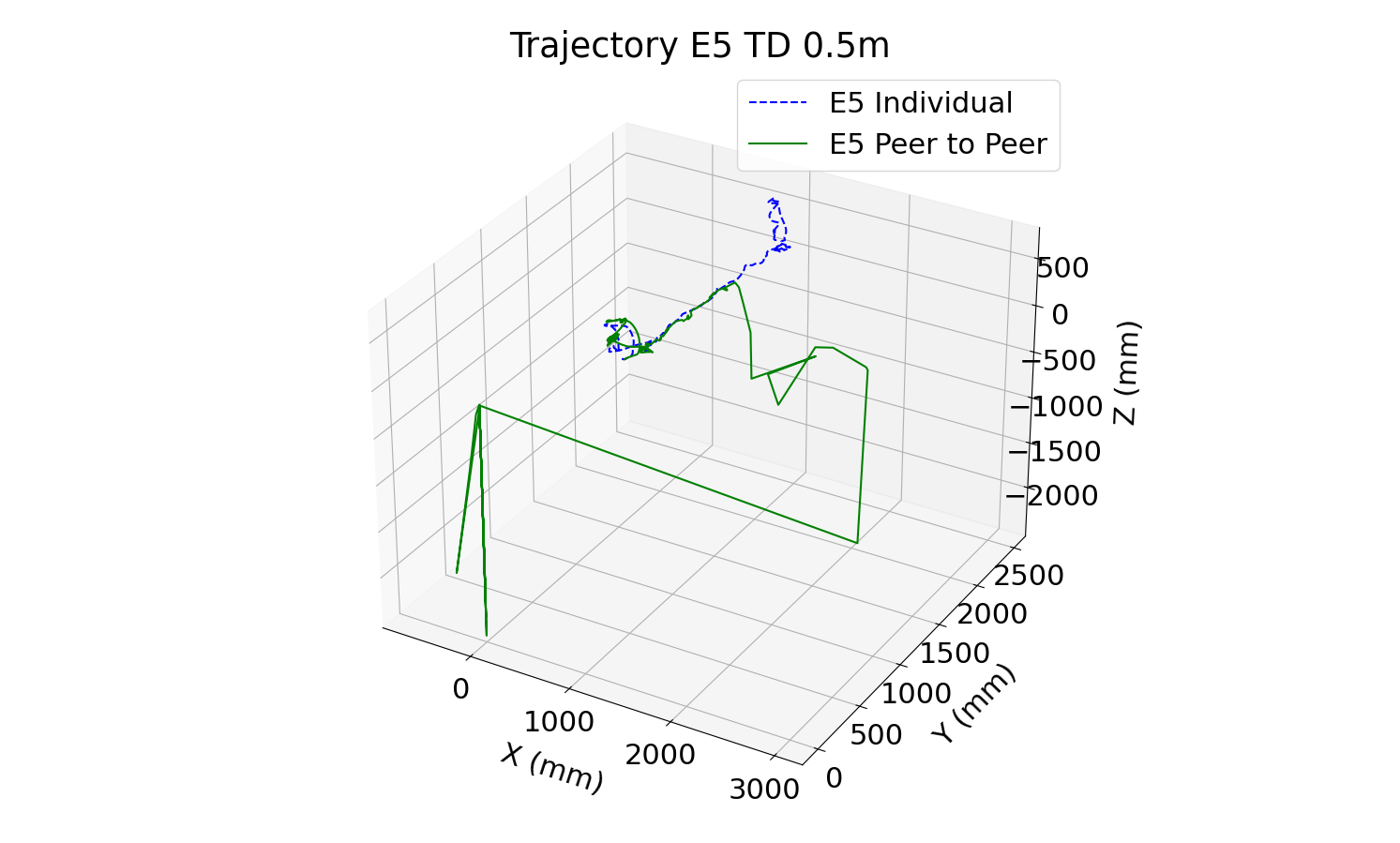}\label{fig:Dynamic-UD-1}}
  \hfill
  \subfloat[Drone E8 - Stable]{\includegraphics[width=0.8\linewidth]{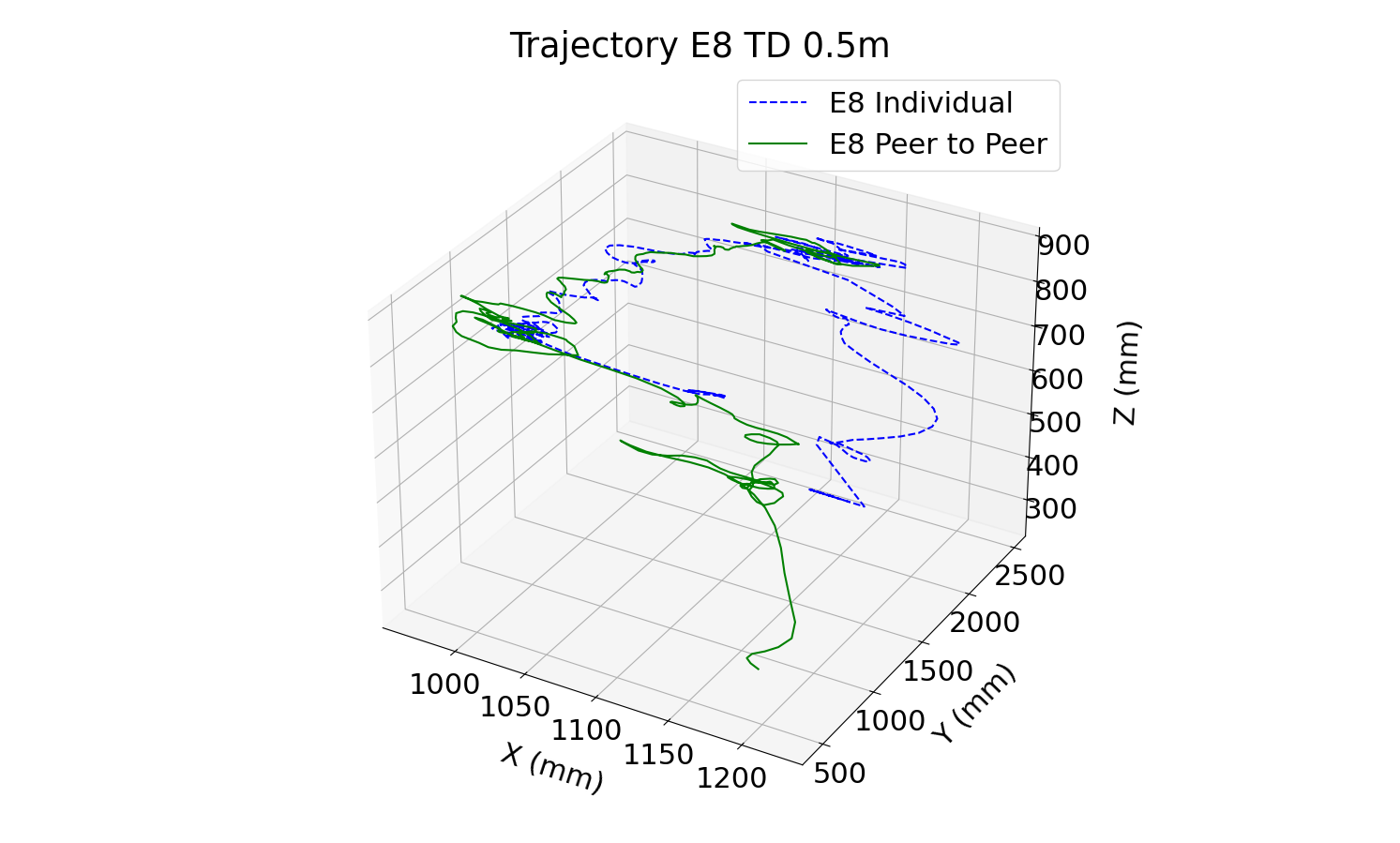}\label{fig:Dynamic-UD-2}}
    \setlength{\abovecaptionskip}{1pt}
    \setlength{\belowcaptionskip}{-3pt}
  \caption{Trajectories of drones in top-down spatial position under high-speed tailwind}
\end{figure}

In side-by-side spatial positioning of drones, drones E5 and E8 consume more energy as they operate in close proximity under high-speed headwinds. As shown in Fig. \ref{fig:Dynamic-SS-1}, drone E8 consumes more energy in the presence of a peer drone (i.e., E5) than when it operates individually. However, drone E8 consumes approximately the same energy when it operates with a lateral separation of 0.6m from the peer drone as it would \emph{individually} (see Fig. \ref{fig:Dynamic-SS-3}).\looseness=-1

\begin{figure}[!t]
\vspace{-20pt}
  \centering
  \subfloat[Drone E5]{\includegraphics[width=0.23\textwidth]{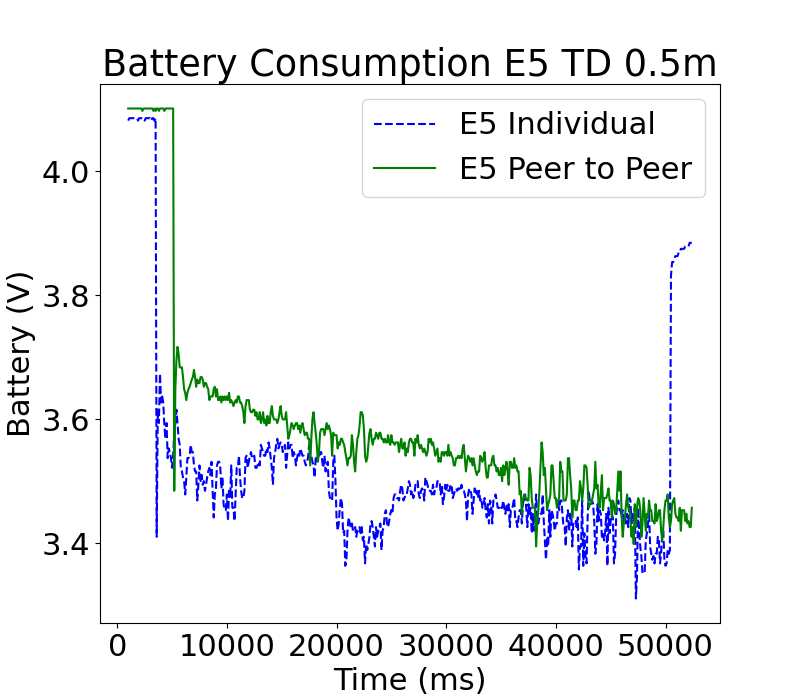}\label{fig:Dynamic-UD-3}}
  \hfill
  \subfloat[Drone E8]{\includegraphics[width=0.23\textwidth]{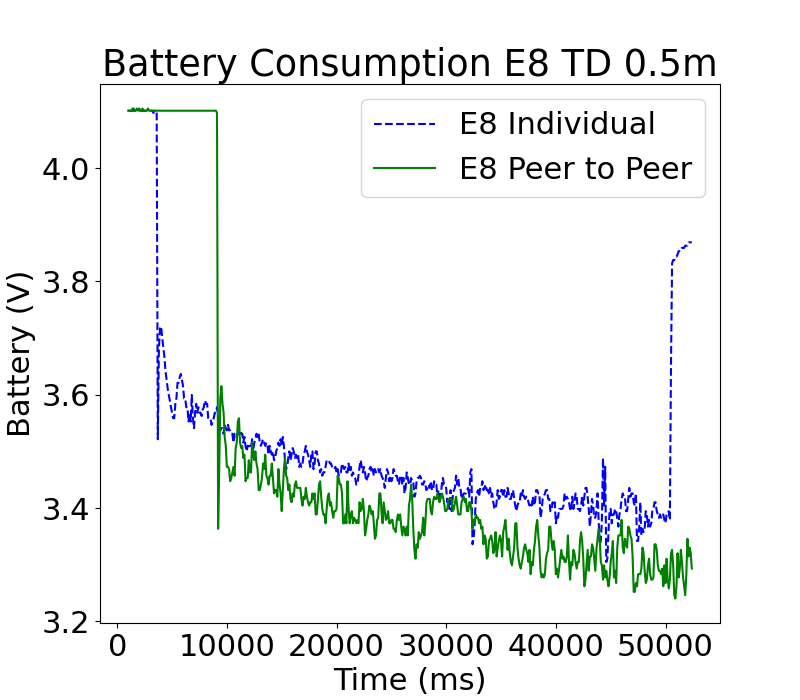}\label{fig:Dynamic-UD-4}}
    \setlength{\abovecaptionskip}{1pt}
    \setlength{\belowcaptionskip}{-10pt}
  \caption{Battery consumption of E5 and E8 operating in top-down under high speed headwind}
\end{figure}

In the top-down spatial position, the drone operating at the lower altitude (i.e., E5) frequently deviated from its planned trajectory (see Fig. \ref{fig:Dynamic-UD-1}). The instability in flight was mainly due to the wind conditions and the impact of the neighboring drone. However, the flights were stable on increasing the inter-drone vertical separation (see Fig. \ref{fig:Dynamic-UD-2}). The drone operating at the lower altitude consumes less battery in peer-drone presence than when operated individually under high-speed headwind conditions (see Fig. \ref{fig:Dynamic-UD-3}). This is due to the external wind that disperses the trailing wind generated by the top drone (i.e., E8) which reduces the impact of the downwash force on the drone operating at lower altitude (i.e., E5). However, there is a slight negative impact on the battery consumption of drone E8 operating at a higher altitude(see Fig.\ref{fig:Dynamic-UD-4}). 

\begin{figure}[h!]
\vspace{-20pt}
  \centering
  \subfloat[Distance of Separation-0.3m]{\includegraphics[width=0.23\textwidth]{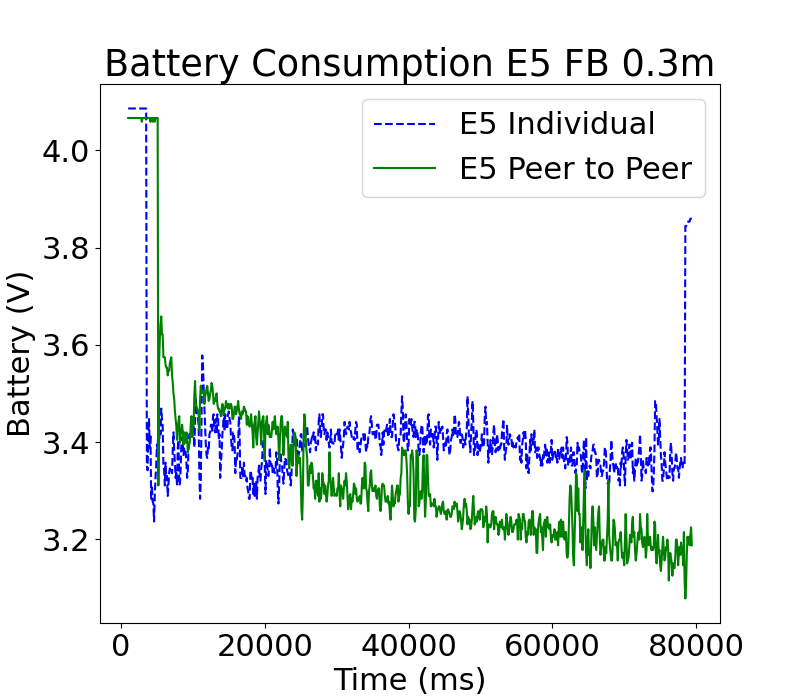}\label{fig:Dynamic-FB-1}}
  \hfill
  \subfloat[Distance of Separation-0.5m]{\includegraphics[width=0.23\textwidth]{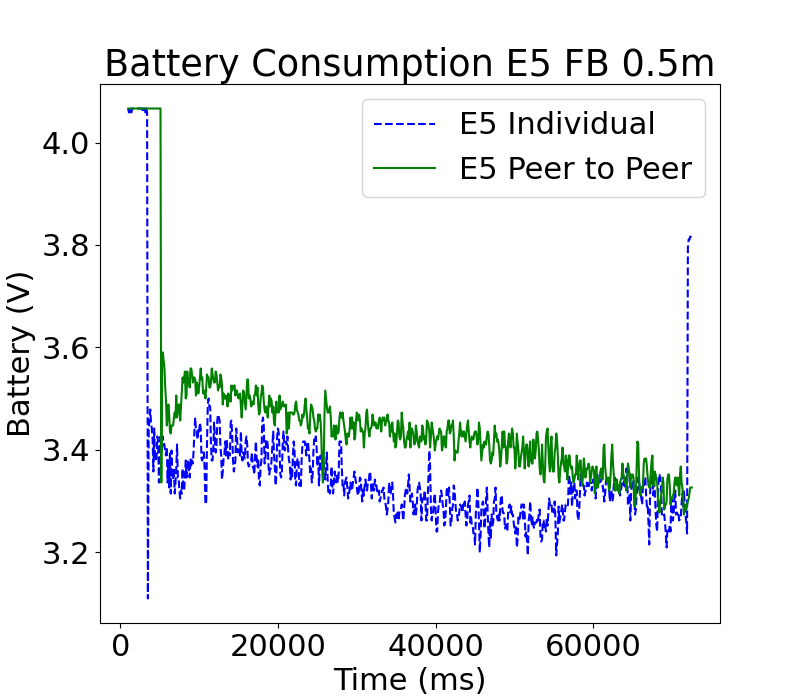}\label{fig:Dynamic-FB-3}}
      \setlength{\abovecaptionskip}{1pt}
    \setlength{\belowcaptionskip}{-10pt}
  \caption{Battery consumption of E5 operating in front-back under high-speed headwind}
\end{figure}

In the front-back peer-to-peer spatial positioning, the impact on battery consumption of drone E5 (i.e., the drone at the back) decreases as the distance of separation with the peer-drone (i.e., E8) increases (see Figs. \ref{fig:Dynamic-FB-1}, \ref{fig:Dynamic-FB-3}). The results are consistent with the results from controlled wind conditions (i.e., no wind) (see Figs. \ref{fig:Static-FB-1}, \ref{fig:Static-FB-3}).\looseness=-1 

In side-by-side offset spatial position, the energy consumption of the trailing drone (i.e., E5) increases as the lateral separation from the leading drone (i.e., E8) increases with an offset along the y-axis under low-speed headwind condition (see Figs. \ref{fig:Dynamic-SSO-1}, \ \ref{fig:Dynamic-SSO-3}). In this case, as the separation distance increases, the impact of upwash forces, which provides lift to drone E5, diminishes. This results in higher energy consumption, which corresponds with findings of \cite{alkouz2020formation}, indicating that drones flying in echelon and vee formations benefit from the upwash trailing forces of the leading drones.\looseness=-1

\begin{figure}[h!]
\vspace{-20pt}
  \centering
  \subfloat[Distance of Separation--0.3m]{\includegraphics[width=0.23\textwidth]{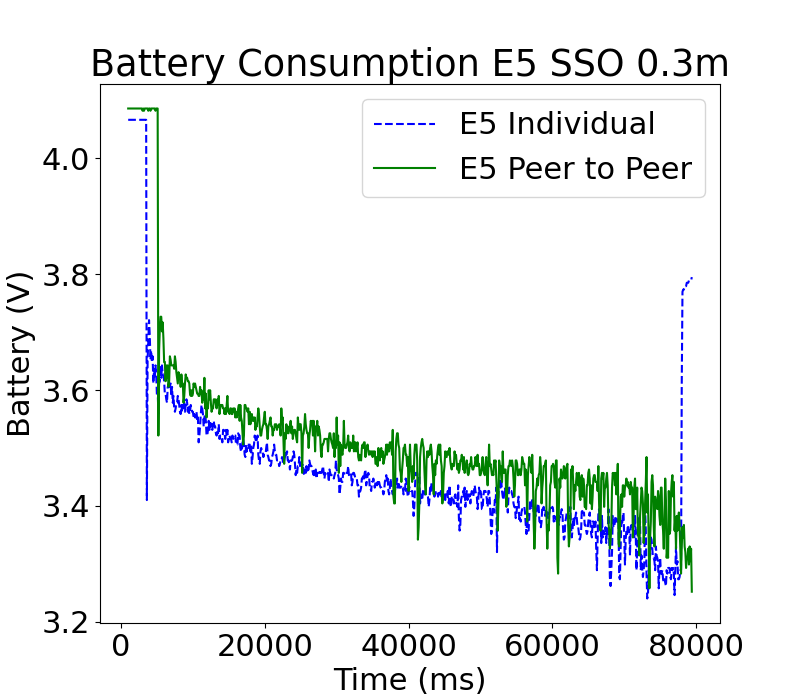}\label{fig:Dynamic-SSO-1}}
  \hfill
  \subfloat[Distance of Separation--0.8m]{\includegraphics[width=0.23\textwidth]{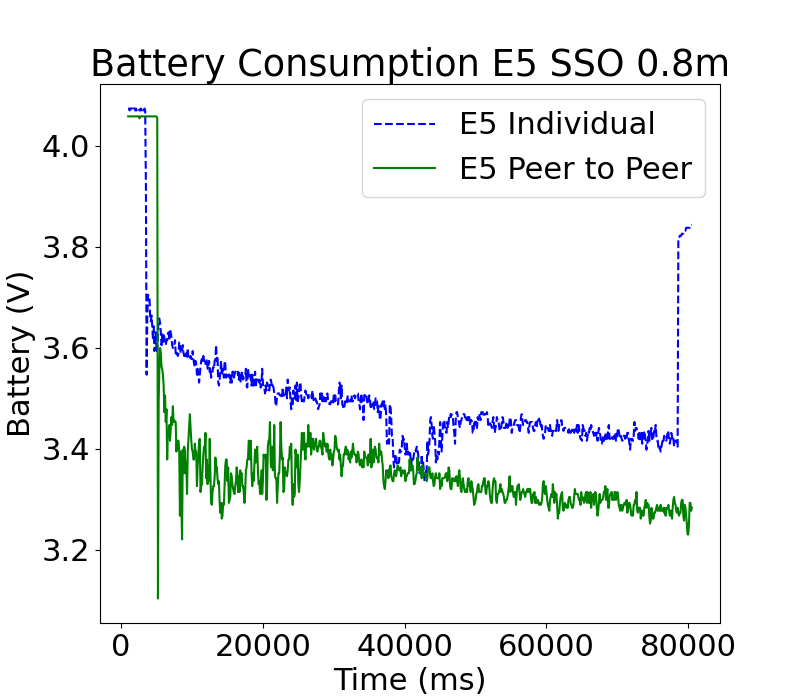}\label{fig:Dynamic-SSO-3}}\setlength{\abovecaptionskip}{1pt}
    \setlength{\belowcaptionskip}{-10pt}
  \caption{Battery consumption of E8 operating in side-by-side offset under low speed head wind}
\end{figure}

\begin{figure}[h!]
\vspace{-20pt}
  \centering
  \subfloat[Distance of Separation--0.3m]{\includegraphics[width=0.23\textwidth]{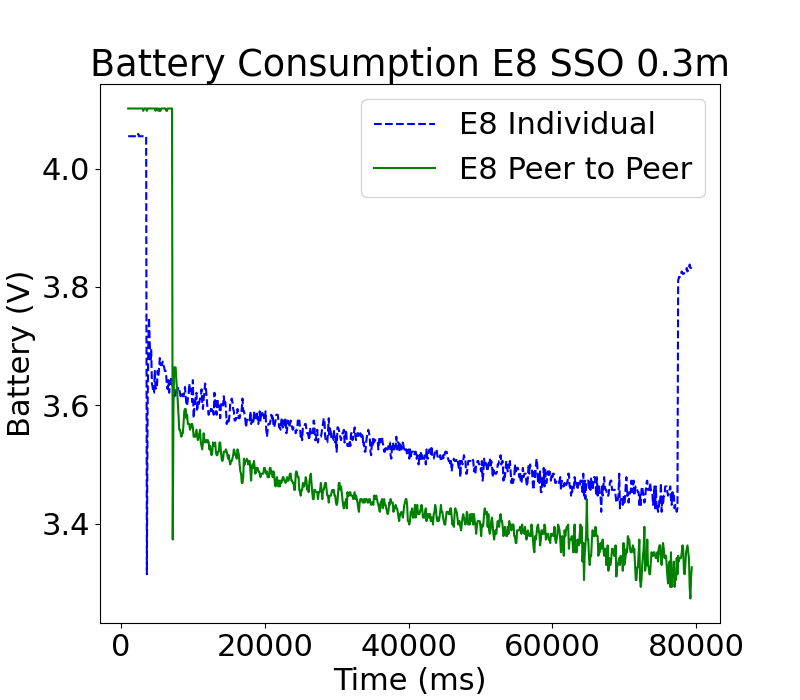}\label{fig:Dynamic-SSO-1}}
  \hfill
  \subfloat[Distance of Separation--0.8m]{\includegraphics[width=0.23\textwidth]{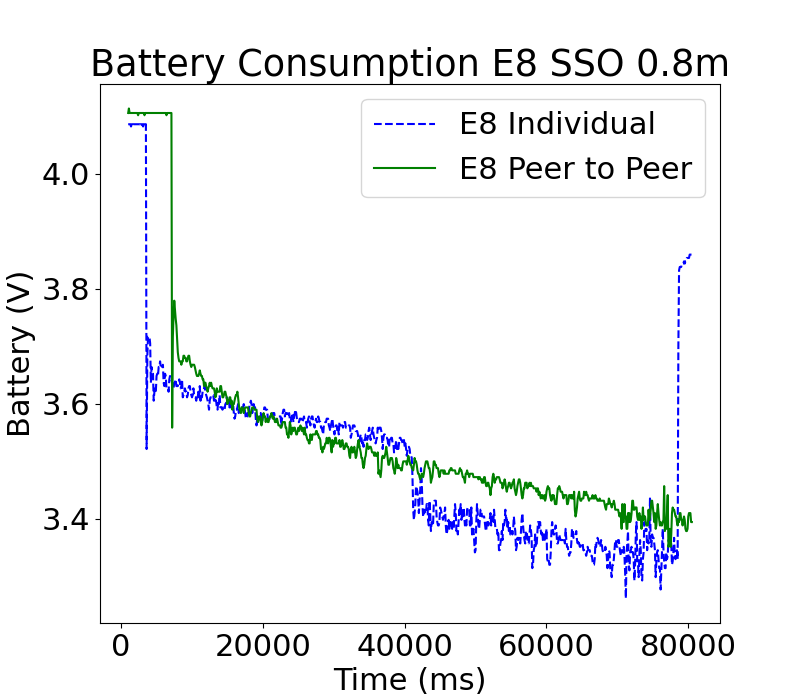}\label{fig:Dynamic-SSO-3}}
    \setlength{\abovecaptionskip}{1pt}
    \setlength{\belowcaptionskip}{-10pt}
    
  \caption{Battery consumption of E5 operating in side-by-side offset under low-speed headwind}
\end{figure}

\begin{figure}[h!]
\vspace{-20pt}
  \centering
  \subfloat[Drone E5]{\includegraphics[width=0.23\textwidth]{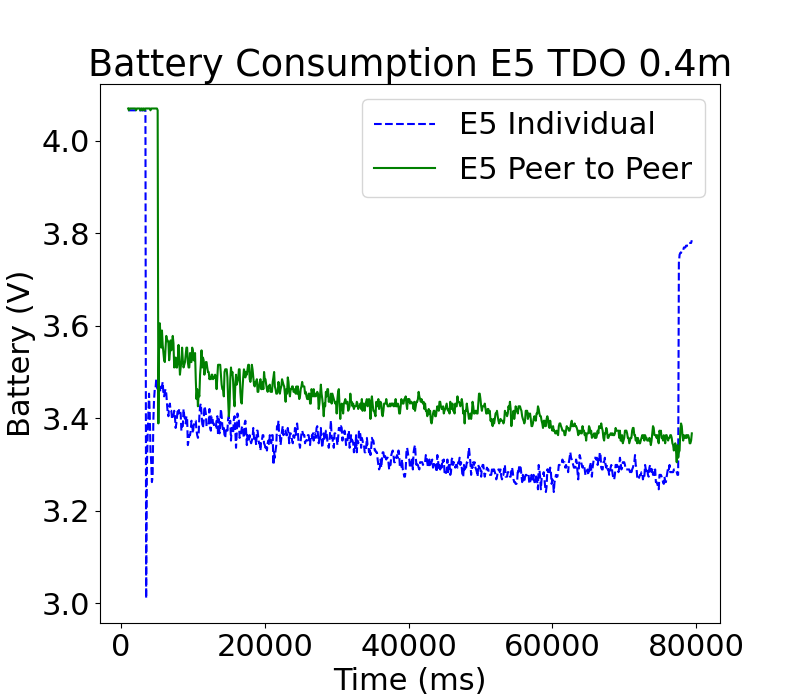}\label{fig:Dynamic-UDO-1}}
  \hfill
  \subfloat[Drone E8]{\includegraphics[width=0.23\textwidth]{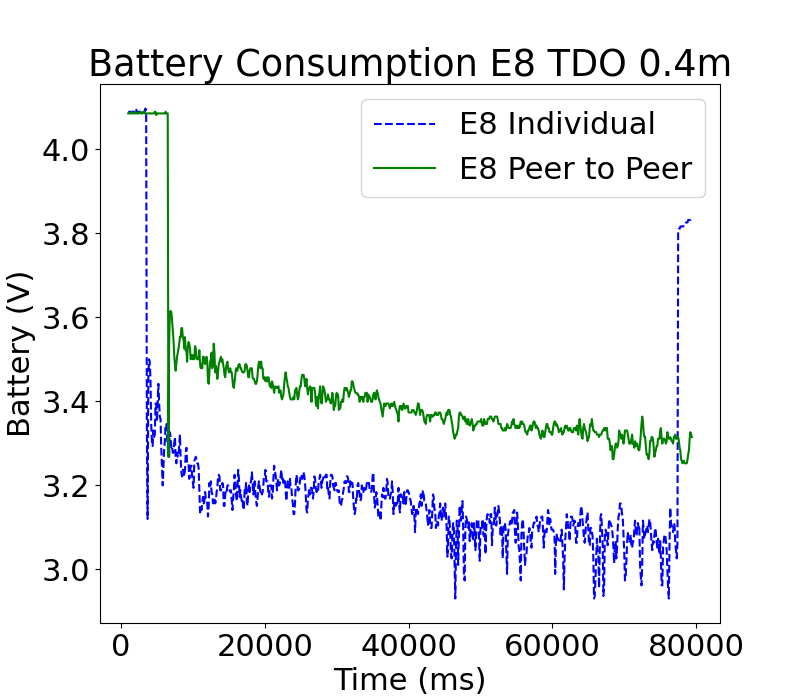}\label{fig:Dynamic-UDO-2}}\setlength{\abovecaptionskip}{1pt}
    \setlength{\belowcaptionskip}{-10pt}
  \caption{Battery consumption of E5 and E8  in top-down with offset and low-speed tailwind}
\end{figure}

In the top-down offset spatial position, the drone operating at a lower altitude (i.e., E5) was positively impacted for both tailwind and headwind (see Fig. \ref{fig:Dynamic-UDO-1}). This is due to the external wind that minimizes the downwash force from the drone E8 operating at higher altitudes. Drone E8 was slightly negatively impacted by headwind (see Fig. \ref{fig:Dynamic-UD-4}) and slightly positively impacted by tailwind (see Fig. \ref{fig:Dynamic-UDO-2}).\looseness=-1 

\vspace{-8pt}
\section{Conclusion}
We demonstrated the impact of spatial proximity on drones' battery consumption under different wind conditions. We conducted a set of experiments for different spatial positioning of drone pairs as they operate in a 3D-space. We designed a GUI to facilitate the visualization and analysis of peer-to-peer impact graphs across different wind conditions. The analysis shows that the battery consumption of drones is impacted when they operate in close proximity to their peer drones compared to when they operate individually. Moreover, the impact depends on the wind condition and positioning of the drone with the peer drone.

\begin{acks}
This research was partly made possible by DP160103595 and LE180100158 grants from the Australian Research Council. The statements made herein are solely the responsibility of the authors
\end{acks}

\bibliographystyle{ACM-Reference-Format}
\balance
\bibliography{main}

\end{document}